\documentstyle[prd,aps,preprint,epsfig,floats]{revtex}
\draft
\tighten

\begin{document}

\preprint{
        \parbox{1.5in}{%
           hep-ph/9801262 \\
           PSU/TH/192
        }
}

\title{Proof of Factorization for Deeply Virtual Compton Scattering in QCD}

\author{John C. Collins and Andreas Freund
}

\address{
        Department of Physics, Penn State University,
        104 Davey Lab., University Park, PA 16802, U.S.A.
}

\date{2 July 1998}

\maketitle

\begin{abstract}

We show that factorization holds for the deeply virtual Compton
scattering amplitude in QCD, up to power suppressed terms, to all
orders in perturbation theory.  The theorem applies to the production
of off-shell photons as well as real photons.  We give a detailed
treatment of the situation where one of the two partons joining the
parton density to the hard scattering has zero longitudinal momentum.

\bigskip


\end{abstract}

%==========================================
\section{Introduction}
\label{sec:intro}

In this paper, we prove factorization for the deeply virtual
Compton scattering (DVCS) amplitude in QCD up to power suppressed
terms, to all orders in perturbation theory. This proof is
important because of the recent great interest in DVCS
\cite{Ji,Rad1,Rad2,muell,Ji1,BM,DGPR,chen,FFS,man}. One important use
of DVCS is as a probe of off-forward (or nondiagonal)
distributions \cite{Ji,Rad2,OFPD,CFS,FFSV}. These differ from
the usual parton distributions probed in inclusive
reactions by having a non-zero momentum transfer between the
proton in the initial and final state.

A related process which is also used to probe off-diagonal parton
densities is exclusive meson production in deep-inelastic
scattering \cite{Brod,Ryskin}, for which a proof of factorization was
given in \cite{CFS}.  Compared with this process, DVCS is simpler
because the composite meson in the final state is replaced by an
elementary particle, the photon, and thus there is no meson wave
function in the factorization formula.

In the case of a scalar field theory, Anikin and Zavialov \cite{AZ}
proved a non-local operator product expansion, from which follows the
factorization theorem for DVCS, as shown by M\"uller et al.\
\cite{MRGDH}.  From the point of view of these papers, the new results
in the present paper consist of an extension of the results to QCD.
However, we do not derive an explicit form of an $R$-operation for the
coefficient functions, unlike Anikin and Zavialov.

In a separate line of development, Ji \cite{Ji} and Radyushkin
\cite{Rad1} provided key insights that
indicate that a factorization theorem is valid for DVCS, and then
Radyushkin provided an all-orders proof in \cite{Rad2}.  In this paper
we provide an alternative proof, and give a new treatment of some
problems that were treated in Ref.\ \cite{Rad2} but that were perhaps not
fully solved.  The primary technical difference between Radyushkin's
derivations and ours is that he uses the $\alpha$-parametric
representation for Feynman graphs, whereas we use the momentum
representation, which we consider to be more direct.
Our proof follows the general lines of proofs of
factorization for other processes given in \cite{CSS,colster}, and the most
noteworthy feature is that, particularly for the case of production of
off-shell photons, the proof is simpler than for any other process.
Even for ordinary deep-inelastic scattering one needs to discuss the
cancellation of soft gluon exchanges and of final-state interactions,
whereas these complications are not present in the leading power for
DVCS.

The paper is organized in the following way: After stating the theorem
in Sec.\ \ref{sec:theorem}, we show in Sec.\ \ref{sec:proof} how the
proof given in Ref.\ \cite{CFS} is readily adapted to DVCS.  The
complications mentioned above concern the situation when one of the
two lines connecting the parton density to the hard scattering carries
zero longitudinal momentum, and these are given a detailed treatment
in Sec.\ \ref{subt}.

%========================================
\section{Factorization Theorem}
\label{sec:theorem}

The process under consideration is DVCS which is the elastic
scattering of virtual photons:
\begin{equation}
\gamma ^{*}(q) + P(p) \rightarrow \gamma ^{(*)}(q') + P'(p-\Delta )
\label{dvcs}
\end{equation}
where the diffracted proton $P'$ may also be replaced by
a low-mass excited state and the
final-state photon can be either real or time-like. This process is the
hadronic part of $ep \rightarrow e \gamma p'$ for a real photon or of
$ep \rightarrow e\mu ^{+}\mu ^{-}p'$ for a time-like photon.

It is convenient to use light-cone coordinates with respect to the collision
axis\footnote
{We define a vector in light cone coordinates by:
\begin{displaymath}
V^{\mu }=\left (V^{+},V^{-},V_{\perp } \right )=
\left ( \frac {V^{0}+V^{3}}{\sqrt {2}},
\frac {V^{0}-V^{3}}{\sqrt {2}},V^{1,2}\right ).
\end{displaymath}
}.
The momenta in the process then take the form:
\begin{eqnarray}
   p^{\mu } &=& \left ( p^{+},\frac {m^{2}}{2p^{+}},{\bf 0}_{\perp } \right ),
\nonumber\\
   q^{\mu } &\simeq& \left ( -xp^{+},\frac {Q^{2}}{2xp^{+}},{\bf 0}_{\perp }
\right ),
\nonumber\\
   q'^{\mu } &\simeq& \left ( xp^{+}\frac {\Delta ^{2}_{\perp }+\alpha
Q^{2}}{Q^{2}},
      \frac {Q^{2}}{2xp^{+}},{\bf \Delta }_{\perp }
   \right ),
\nonumber\\
   \Delta ^{\mu } &\simeq& \left ( x(1+\alpha )p^{+},
      -\frac {\Delta ^{2}_{\perp } + m^{2} (1+\alpha ) x}{2(1-x-\alpha
x)p^{+}},
      {\bf \Delta }_{\perp } \right ).
\label{momenta}
\end{eqnarray}
Here, $x$ is the Bjorken scaling variable,
$Q^{2}$ is the virtuality of the initial photon,
$m^{2}$ is the proton mass, $t=\Delta ^{2}$ is
the momentum transfer squared, and
$\alpha $ is a parameter that specifies the virtuality of the
outgoing photon: ${q'}^{2} = \alpha Q^{2}$.  Thus, $\alpha =0$ for a
real photon and $\alpha >0$ for a time-like photon. Finally,
$\simeq$ means ``equality up to power suppressed terms''.

The theorem to be proved is that the amplitude for the process (\ref{dvcs})
is:
\begin{equation}
T = \sum_{i} \int_{-1+x}^{1} dx_{1} f_{i/p}\left ( x_{1},x_{2},t,\mu \right )
H_{i}\left (x_{1}/x,x_{2}/x,\mu \right )\, + \, 
\mbox{power-suppressed corrections},
\label{theorem}
\end{equation}
where $f_{i/p}$ is a nondiagonal parton distribution and
$H_{i}$ is the hard-scattering coefficient for scattering off a
parton of type $i$.
We let $x_{1}$ be the momentum fraction of parton $i$ coming from
the proton, so that $x_{2}=x_{1} - x(1+\alpha)$ is
the momentum fraction which is 
returned to the proton by the other parton line joining the parton
distribution and the hard scattering.  There is implicit
polarization dependence in the amplitude.
$\mu $ is the usual
renormalization/factorization scale, which should be of order $Q$
to allow
calculations of the hard scattering coefficients
within finite-order perturbation
theory. The $\mu $ dependence of $f_{i/p}$ is given by equations of the
DGLAP and Brodsky-Lepage
kind \cite{Ji,Rad1,Rad2,OFPD,Brod,CFS,FFSV}.
The parton distributions in Eq.\ (\ref{theorem}), together with
their evolution equations, are defined using the conventions of
\cite{CFS,FFSV}.  They may easily be transformed into those given
in \cite{Ji,Rad1,Rad2} by a change of normalization and of kinematic
variables.

%========================================
\section{Proof of Theorem}
\label{sec:proof}

The proof of our theorem Eq.\ (\ref{theorem}) can be summarized as follows
\footnote{
For a very detailed account of the basic steps and potential problems
see Ref.\ \cite{CFS}.
}:
\begin{itemize}

\item Establish the non-ultra-violet regions in the space of loop momenta
      contributing to the amplitude.

\item Establish and prove a power counting formula for these regions.

\item Determine the leading regions of the amplitude.

\item Define the necessary subtractions in the amplitude to avoid double
      counting.

\item Taylor expand the amplitude to obtain a factorized form.

\item Show that the part containing the long-distance information can be
      expressed through matrix elements of renormalized, bi-local, gauge
      invariant operators of twist-$2$.

\end{itemize}

%----------------------
\subsection{Regions}
\label{reg}

First let us establish the regions in the space of loop momenta contributing
to the asymptotics of the amplitude, i.e.,\ the generalized reduced graphs.
The steps leading to the generalized reduced graphs are identical to the steps
1--3 in Sec.\ IV of Ref.\ \cite{CFS}, i.e.,\ scale all momenta by a factor
$Q/m$, use the Coleman-Norton theorem to locate all pinch-singular surfaces
in the space of loop momenta (in the zero-mass limit), and finally identify
the relevant regions of integration as neighborhoods of these pinch-singular
surfaces.

\begin{figure}
\centering
\mbox{\epsfig{file=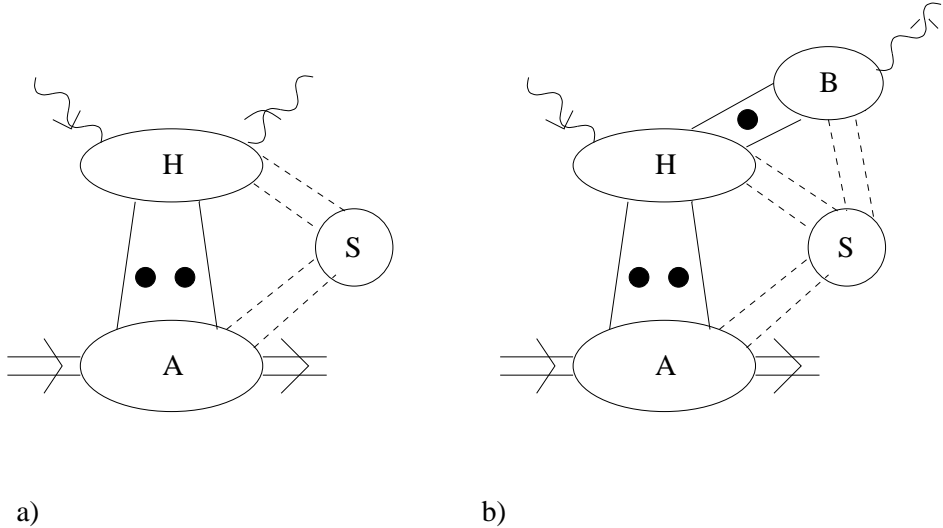,height=7cm}}
\vspace*{5mm}
\caption{a) Reduced graph for DVCS with direct coupling for
the out-going photon to hard subgraph.
b) The same without a direct like coupling for the
out-going photon. }
\label{Reduced.Graphs}
\end{figure}

The results are the two kinds of reduced graph shown in Fig.\
\ref{Reduced.Graphs}.  There, $A$ and $B$ denote collinear graphs
with one large momentum component in the $+$ and $-$ direction
respectively, $H$ denotes the hard scattering graph, and $S$
denotes a
graph with all of its lines soft, i.e.,\ in the center-of-mass
frame all the components of the momenta in $S$ are much smaller than
$Q$.
Note that, of the external momenta, $p$ and $p'$ belong to $A$,
$q'$ belongs to $B$ or $H$ , and $q$ belongs to $H$.

When the two external photons have comparably large virtualities,
the only reduced graphs are of the first kind, Fig.\
\ref{Reduced.Graphs}a, where the out-going photon couples
directly to the hard scattering. But when the out-going photon
has much lower virtuality than the incoming photon, for example,
when it is real, we can also have the second kind of reduced
graph, Fig.\ \ref{Reduced.Graphs}b, where the out-going photon
couples to a $B$ subgraph. As we will see later, power counting
will show that the second kind of reduced graph, Fig.\
\ref{Reduced.Graphs}b, is power suppressed compared to the first
kind, with a direct photon coupling. This implies that we will
avoid all the complications which were encountered in \cite{CFS}
that are associated with the meson wave function.

%----------------------
\subsection{Power Counting}
\label{count}

Each reduced graph codes a region of loop-momentum space, a
neighborhood of the surface $\pi $ of a pinch singularity in the
zero-mass limit.
The contribution to the amplitude from a neighborhood of $\pi $
behaves like $Q^{p(\pi )}$, modulo logarithms, in the large-$Q$
limit, with the power given by
\begin{eqnarray}
    p(\pi ) &=& 4 - n(H) - \mbox{\#(quarks from $S$ to $A$, $B$)} -
    3\mbox{\#(quarks from $S$ to $H$)}\nonumber\\
    & & - 2\mbox{\#(gluons from $S$ to $H$)}.
\label{pform}
\end{eqnarray}
where $n(H)$ is the number of collinear quarks, transversely
polarized gluons,
and external photons attaching to the hard subgraph $H$.
Such results were obtained by Libby and Sterman
\cite{LibSt,Sterman}.
The particular form of Eq.\ (\ref{pform}) was given in \cite{CFS}
together with a proof that applies without change to DVCS.

The well-known problem of gluons with scalar polarization (see,
for example, \cite{CSS,ER,LabSt}) will be dealt with later on.  Suffice it
to say here that gluons with such a polarization can be
factorized into the parton distributions by using
gauge-invariance arguments.

\begin{figure}
\centering
\mbox{\epsfig{file=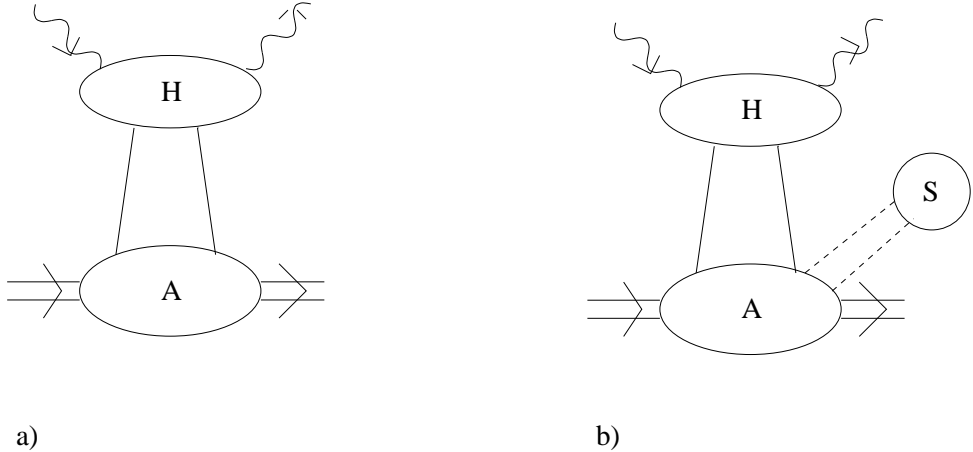,height=6cm}}
\vspace*{5mm}
\caption{Those reduced graphs that contribute to the leading
regions in DVCS. }
\label{Leading.Regions}
\end{figure}

%----------------------
\subsection{Leading Regions}
\label{lreg}

The leading regions for the amplitude are those with the largest
exponent $p(\pi )$ in Eq.\ (\ref{pform}).  It is easy to see that
these correspond to the reduced graphs in Fig.\
\ref{Leading.Regions}, independently of whether the out-going
photon is real or far off-shell.  The corresponding power is $Q^{0}$.
These reduced graphs have direct photon couplings to the hard
subgraph, they have exactly two parton lines connecting the
collinear subgraph $A$ to the hard subgraph $H$, and they have no
soft lines connecting to $H$. The two kinds of graph differ only
by the absence or presence of a soft subgraph that connects to
$A$ alone.

Among the other reduced graphs, which are non-leading for our
process, are those of the type
in Fig.\ \ref{Reduced.Graphs}b, which are leading in the case of
diffractive meson production, where the leading region gives
$Q^{-1}$.

In the case of a photon that is off-shell by order $Q^{2}$,
the amplitude for production
of the photon behaves like $Q^{0}$, the same as for a real photon.
However, the physically observed process includes the decay of
the time-like photon (to a $\mu ^{+}\mu ^{-}$, for example), which
results in a power suppression of the observed cross section by
$1/Q^{2}$ compared with the cross section for making real photons.

%----------------------
\subsection{Proof of absence of a soft part in leading regions}
\label{poaos}

As mentioned in Sec.\ \ref{lreg}, there might, in principle,
be a soft part $S$ in the leading reduced graph connected solely
to the $A$ graph by gluons, as shown in Fig.\
\ref{Leading.Regions}b. Note that by Eq.\ (\ref{pform}), quarks
connecting $S$ to $A$ would lead to a power
suppression. We will now show that this soft part $S$ is indeed
absent, and so we only need to consider regions of the form of
Fig.\ \ref{Leading.Regions}a.

\begin{figure}
\centering
\vspace*{0.5cm}
\mbox{\epsfig{file=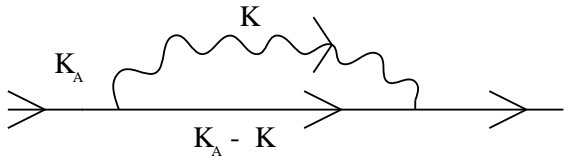,height=2cm}}
\vspace*{0.5cm}
\caption{Soft gluon loop attaching to collinear line.}
\label{Soft.to.A}
\end{figure}

We will first examine a simple one loop example, Fig.\
\ref{Soft.to.A}.  The external quark is part of the $A$ subgraph
in Fig.\ \ref{Leading.Regions}b, and the gluon is soft. So we
parameterize the momenta by:
\begin{eqnarray}
   k_{A} &=& (x_{1}p^{+}, k_{A}^{-}, {\bf k_{A,\perp }}),
\nonumber\\
   k  &=& (k^{+},k^{-},{\bf k_{\perp }}).
\label{mom}
\end{eqnarray}
where the $k^{+}_{A}$ is $O(Q)$ and all the other components are of
$O(m)$ or smaller.

If we omit irrelevant factors in the numerator,
the loop integral takes the following form:
\begin{eqnarray}
& &\int _{{\rm soft}~k} d^{4}k
  \frac {1}{(k^{2}+i\epsilon )\left [(k_{A}-k)^{2} -m^{2} +i\epsilon \right ]}
\nonumber\\
& &~~=\int _{{\rm soft}~k} d^{4}k
  \frac {1}{(2k^{+}k^{-} -k_{\perp }^{2}+i\epsilon ) \left [2(x_{1}p^{+} -
k^{+}) (k_{A}^{-}-k^{-})-(k_{A,\perp }-k_{\perp })^{2}-m^{2} +i\epsilon \right
]}
\nonumber\\
& &~~\simeq \int _{{\rm soft}~k} dk_{+}
  \frac {1}{(2k^{+}k^{-} -k_{\perp }^{2}+i\epsilon ) \left [2x_{1}p^{+}
(k_{A}^{-}-k^{-})-(k_{A,\perp }-k_{\perp })^{2}-m^{2} +i\epsilon \right ]}.
\label{softk}
\end{eqnarray}
As before, $\simeq$ stands for ``equality up to power suppressed terms''.
As one can see, there is no $k^{+}$-pole in the second part of the denominator
and we can freely deform the contour in $k^{+}$ to avoid the pole in the soft
gluon propagator.  This takes us out of the soft region for $k$.

In the general situation, Fig.\ \ref{Leading.Regions}b, we can
use a version of the arguments in Ref.\ \cite{CFS,C97} to show
that the soft momenta ${k^{+}_{i}}$ can be rerouted in such a way
as to exhibit a lack of a pinch singularity.  The essential idea
is that one can find a path backwards or forward from one
external line of $S$ to another external line of $S$.  The loop
is completed along lines of $A$, all of which have much larger
$+$ momenta than what is typical of soft momenta, and hence there
is no pinch.

%----------------------
\subsection{Subtractions}
\label{sub}

The subtractions necessary to avoid double counting in the amplitude are
treated exactly the same fashion as the ones in Sec.\ VI of Ref.\ \cite{CFS}, 
since the distributional arguments to construct the subtraction terms on a
pinch-singular surface $\pi $ presented there are very general in nature and
are not limited to the case of diffractive vector meson
production that was
considered in \cite{CFS}.

The above statement leads to the following asymptotic form of the amplitude
\begin{equation}
\mbox{Asy}\, T = \Sigma_{\Gamma }\mbox{Asy}\, \Gamma = A \times H.
\end{equation}
where $\Gamma $ stands for a possible graph for the amplitude $T$.

%----------------------
\subsection{Taylor expansion}
\label{TE}

We now obtain the leading term in the hard subgraph, when
it is expanded in powers of the small momenta.
The arguments used are exactly analogous to the ones used
in Sec.\ VII of Ref.\ \cite{CFS} except that we do not have to deal with a
$B$ subgraph as was the case in \cite{CFS}. So we have:
\begin{eqnarray}
 A \times H &\simeq&
    \int dk_{A}^{+}
    \,
    H\left(q,q',(k_{A}^{+},0,0_{\perp }),
    (\Delta ^{+} - k_{A}^{+},0,0_{\perp }
    \right)
\nonumber\\
&&
    \int dk_{A}^{-}d^{2}k_{A,\perp }
    \,
    A(k_{A},\Delta -k_{A}),
\label{CTE}
\end{eqnarray}
where $k_{A}$ is the loop momentum joining the $A$ and $H$
subgraphs,
and again $\simeq$ means equality up to power-suppressed corrections.
Eq.\ (\ref{CTE}) has
already a factorized form.
However we still have to deal with the extra scalar gluons that
may be exchanged between the subgraphs $A$ and $H$;
this will be done in the next subsection.

Eq.\ (\ref{CTE}) can be written in the following way:
\begin{equation}
A \times H \simeq \Sigma_{i} \int dk^{+}
C_{i}(q,q',k^{+})O_{i}(p,p',k^{+}),
\label{sumeq}
\end{equation}
where the $C_{i}$ are the short distance coefficient functions and the $O_{i}$
are the matrix elements of renormalized light-cone operators.

%----------------------
\subsection{Gauge Invariance}
\label{gi}

\begin{figure}
\centering
\mbox{\epsfig{file=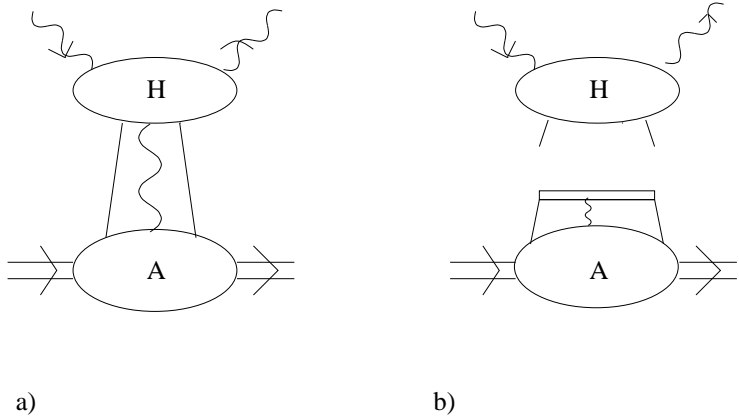,height=5.5cm}}
\vspace*{5mm}
\caption{a) A scalar gluon attaching the collinear subgraph to
the hard subgraph $H$ in the unfactorized form. b) Factorized form after
application of gauge invariance and Ward-identities. The double line
represents the eikonal line to which the scalar gluon attaches.}
\label{Scalar.Gluon}
\end{figure}

In order to identify the $O_{i}$ with the parton distributions as
defined in \cite{CFS} (for example), it is necessary to show that all
gluons with scalar polarization attaching to the hard graph can be
combined into a path-ordered exponential.  Fig.\ \ref{Scalar.Gluon}
shows the example of one scalar gluon.  This was shown in
\cite{CSS,ER,LabSt}.  Another version of a proof was given in Sec.\
VII.\ D of Ref.\ \cite{CFS}, which relies on very general results
recently obtained by Collins \cite{col}.
In this way we obtain exactly the same parton distributions as in \cite{CFS},
namely:
\begin{eqnarray}
   f_{q/p} &=& \int ^{\infty }_{-\infty }\frac {dy^{-}}{4\pi
}e^{-ix_{2}p^{+}y^{-}}
      \langle p| T\bar \psi (0,y^{-},{\bf 0_{\perp }})\gamma ^{+}{\cal P}\psi
(0)|p'\rangle ,
\nonumber\\
   f_{g/p} &=& -\int ^{\infty }_{-\infty }\frac {dy^{-}}{2\pi }\frac
{1}{x_{1}x_{2}p^{+}}
   e^{-ix_{2}p^{+}y^{-}}
   \langle p| T G_{\nu }^{+}(0,y^{-},{\bf 0_{\perp }}){\cal P}G^{\nu
+}(0)|p'\rangle .
\end{eqnarray}
Here, $\cal P$ represents a path-ordered exponential of the gluon
field that makes the operators gauge invariant.  The variable
$x_{2}$ is the same as in Eq.\ (\ref{theorem}). The evolution
equations are the same as in \cite{Rad1,Rad2} and \cite{CFS}.

%----------------------
\subsection{Partons with $k^+=0$: breakpoints and endpoints}
\label{subt}

In the factorization theorem Eq.\ (\ref{theorem}), the integral over
the fractional momenta includes the points $x_1=0$ and $x_2 = 0$.  At
these points, the hard scattering coefficient for DVCS has a pole, and
so we appear to get a logarithmic contribution to the cross section
from a region in which one of the lines joining the parton density to
the hard scattering subgraphs is soft instead of collinear.  This
apparently contradicts our power-counting result that such a region
gives a non-leading power.  This phenomenon was investigated by
Radyushkin \cite{Rad2}.  In this section, we will use momentum-space
methods to give a general
demonstration that the region in question does not give a problem.

First, let us observe that the region of integration over $x_1$ in the
factorization formula Eq.\ (\ref{theorem}) is $-1+x\leq x_1 \leq 1$.
This is proved by the methods of light-front perturbation theory, by
requiring that the intermediate states in Fig.\ \ref{Leading.Regions}
be physically allowed.  See Ref.\ \cite{Rad2,FFSV} for detailed
derivations and discussions.  The points $x_1=0$ and $x_2=0$ at which
the potential problem arises are what we will call ``breakpoints'',
since they occur in the middle of the range of integration where one
of the two lines changes direction.

\begin{figure}
\centering
\vspace*{0.5cm}
\mbox{\epsfig{file=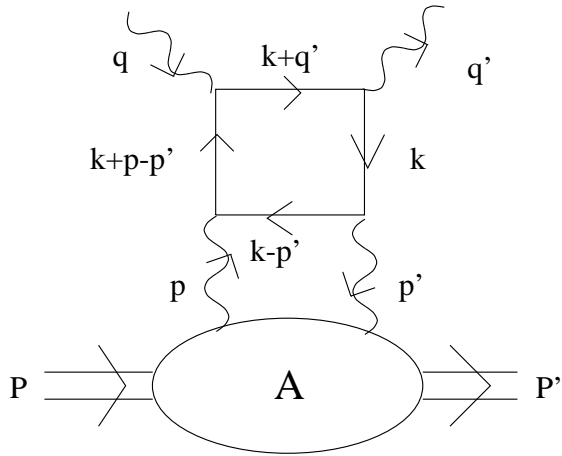,height=6cm}}
\vspace*{0.5cm}
\caption{Particular example of potentially problematic diagram.}
\label{problems}
\end{figure}

We continue by examining a particular case, illustrated in Fig.\
\ref{problems}, and showing how the argument generalizes.  To simplify
the example, let us restrict our attention to regions where the
subgraph $A$ and the lower three lines ($p$, $p'$ and $k-p'$) have
their momenta collinear to the proton. We will also require the two
quark lines, $k+p-p'$ and $k$, on the sides of the ladder to have
their momenta either collinear to the proton or soft.
The example is very similar to one treated by Radyushkin in Ref.\ 
\cite{Rad2}.

We will also only need the case of the production of a real photon,
${q'}^2 = 0$, since this is where the problem arises.

The top loop of the graph has the form:
\begin{equation}
U = \int d^4k
   \frac{{\rm Numerator~factors}}{
       \left( k^2 - m^2 + i\epsilon \right) 
       \left[ (k-p')^2 - m^2 + i\epsilon \right]
       \left[ (k+p-p')^2 - m^2 + i\epsilon \right]
       \left[ (k+q')^2 - m^2 + i\epsilon \right]
} .
\label{probint}
\end{equation}
When both $k$ and $k+p-p'$ are collinear to $A$, the top line is
off-shell by $O(Q^2)$, and it is correct to use the collinear
approximation
\begin{equation}
    \frac{1}{(k+q')^2 - m^2 + i\epsilon}
  \to 
    \frac{1}{ x_2 Q^2 / x + i\epsilon} ,
\label{coll.approx}
\end{equation}
where $x_2 = k^+/p^+$.
A corresponding replacement is also to be
made in the numerator in $U$. The right-hand side of Eq.\
(\ref{coll.approx}) exhibits the afore-mentioned pole at $x_2=0$.
The result of applying the collinear approximation is to give the
appropriate contribution to the factorization formula Eq.\
(\ref{theorem}). 

The collinear approximation becomes invalid when $k$ becomes soft,
i.e., when $x_2 \to 0$.  We must now demonstrate two facts.  The first
is that, when $k$ is in a neighborhood of the soft region, the
collinear approximation is valid after integration over $k$.  The
second fact is that the use of the collinear approximation does not
give an important contribution from some other region of $k$ 
which is absent in the original, unapproximated graph.
We now examine the integral $U$ in the neighborhood of the soft region
for $k$.  It has the following form:
\begin{eqnarray}
U_{{\rm soft}~k} &\simeq& \int_{{\rm soft}~k} d^4k
\frac{{\rm Numerator~factors}}
{ \left[ 2k^{+}k^{-}-k^2_{\perp} - m^2 + i\epsilon \right]
  \left[ 2{p'}^{+}({p'}^- - k^{-}) - (p'-k)^2_{\perp} 
         - m^2 + i\epsilon \right]
}
\label{U.soft}
\\
&&\frac{1}
{\left[ 2 (p^{+}-{p'}^+) (p^- -{p'}^- +k^{-})
 - (p-p'+k)^2_{\perp} - m^2 + i\epsilon \right]
 \left( \frac{k^{+}Q^2}{xp^{+}}-k^2_{\perp} - m^2 + i\epsilon \right)
} ,
\nonumber
\end{eqnarray}
where we have neglected $k^{+}$ in the collinear-to-$A$ lines, $k^{-}$
in the collinear-to-$B$ line and $x\neq x_1$. We have also ignored the
numerator factors, which are an irrelevant complication for our
purposes.

According to the power-counting results of Sec.\ \ref{count}, which
are obtained from \cite{C97}, the soft region for $k$ gives a
power-suppressed contribution.  This estimate assumes that all
components of $k$ are comparable (in the Breit frame), and is obtained
as follows.  Let the magnitude of the components of $k$ be $m$.
Then the order of magnitude of the
soft part of $U$ is a product of factors 
$1/(m^5Q^3)$ from the denominators, $m^4$ from the phase space, and
$Q^2m^2$ from the numerator, for an overall power $m/Q$.
This result can be obtained by writing down the largest 
components in the trace and propagators of Fig.\ \ref{problems} and
Eq.\ (\ref{U.soft}).
Moreover in this region it is
correct to replace the fourth propagator in Eq.\ (\ref{U.soft}) by its
collinear approximation Eq.\ (\ref{coll.approx}), so that we do not
lose the factorization theorem.

However, if the components of $k$ are asymmetric, this estimate no
longer holds.  In particular if the longitudinal components of $k$ are
of order $k^\pm \sim m^2/Q$ while the transverse components remain of
order $m$, then we get contributions of order 
$1/m^8$ from the denominators, $m^6/Q^2$ from the phase space, and
$Q^2m^2$ from the numerators, for a total of $m^0Q^0$. 
This shows that the contribution from this region
is unsuppressed for large $Q$.

At this point we must appeal to the contour deformation arguments of
Ref.\ \cite{C97}.  It is only when the integration over $k$ is pinched
in the region in question that it needs to be taken into account.  In
the dangerous region we have $k^+k^- \ll k_\perp^2$, so that the only
$k^+$ dependence in Eq.\ (\ref{U.soft}) is the pole in the fourth
denominator.  We can therefore deform $k^+$ into the complex plain a
long way out of the region we are considering, indeed all the way to
the collinear-to-$A$ region.  Then the collinear approximation is
valid so that we can replace the graph by its contribution to the
factorization formula.

This contour deformation argument is completely general, as explained
in Sec.\ IIIE of Ref.\ \cite{C97}.  Whenever we have a soft momentum
with $k^+k^- \ll k_\perp^2$, the contour of $k^+$ can be deformed away
from poles in the jet subgraph associated with with the produced
photon.  Since all the relevant singularities are in the final state,
they are all on the same side of the real axis.

Now that we have established in more detail that the only leading
regions are those symbolized in Fig.\ \ref{Leading.Regions}(a), we can
apply the collinear approximation as described earlier, and hence we
obtain the factorization theorem.  

But we still see the following problem.  In the factorization theorem,
Eq.\ (\ref{theorem}), the parton densities are non-analytic at the
breakpoints $x_1=0$ and $x_2=0$, whereas the coefficient function has
a pole at each of these points.  Again consider the collinear
approximation to Fig.\ \ref{problems} in the region we were
considering.  The parton density is non-analytic when $x_2=0$, while
the coefficient function has a pole there, as is seen from the
right-hand-side of Eq.\ (\ref{coll.approx}).  So we cannot literally
apply the contour deformation argument.  

What we will show is that the parton density is continuous\footnote{
   Radyushkin indicated briefly in Ref.\ \cite{Rad2} how such a property
   is to be proved from the $\alpha$ representation for his double
   distributions.
}
at the
breakpoint, so that it can be written as the sum of a function that is
analytic at $x_2=0$ and a function that has a zero at $x_2=0$.  The
only potential leading twist contribution near the breakpoint is
associated with the non-zero analytic term to which the contour
deformation argument applies.

\begin{figure}
\centering
\vspace*{0.5cm}
\mbox{\epsfig{file=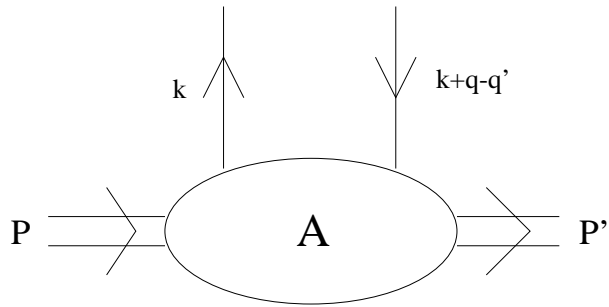,height=4cm}}
\vspace*{0.5cm}
\caption{Parton distribution amplitude.}
\label{pargra}
\end{figure}

To prove this property of the parton density at a breakpoint, consider
a general graph for the parton density, as shown in Fig.\
\ref{pargra}. We have found it convenient to change the labeling of
the momentum compared with the previous figure.
As always, the $k^{-}$ and ${\bf k_{\perp}}$ components of $k$
have been short circuited and are integrated over.
The $k$-line gives a pole at
$k^{-} = (k^{2}_{\perp} + m^2 - i\epsilon) / 2k^{+}
       = (k^{2}_{\perp} + m^2 - i\epsilon) / 2 x_1 P^{+}$, 
while the $k+q'-q$-line gives a pole at
$k^{-} = [(k+q-q')^{2}_{\perp} + m^2 - i\epsilon] / 2 (k^{+}-\xi P^{+})
       = [(k+q-q')^{2}_{\perp} + m^2 - i\epsilon] / 2 x_2 P^{+}$.
Here, $\xi$ is the fractional longitudinal momentum transfer
$1-{P'}^+/P^+$. In addition there are poles from the collinear-to-$A$
in the blob.  For example if the blob consists of a single line, we
have a pole at 
$k^{-} = P^- - (k^{2}+m^2_{\perp}-i\epsilon) / 2 (1-x_1) P^{+}$
or at
$k^{-} = - {P'}^- + [(k+P')^{2}+m^2_{\perp}-i\epsilon) / 2 (1-\xi+x_1)P^{+}$. 

As we vary $x_1$, the $k^-$ contour can generally be deformed to avoid
the poles, so that we have analytic dependence on $x_1$.  The possible
exceptions occur when the $k^-$ contour is pinched for finite $k^-$ or
when a singularity coincides with the endpoint of the integration at
$k^-=\infty$.  A pinch never occurs; in the general case this is a
consequence of the Landau rules.  But endpoint singularities occur,
and these are precisely at the breakpoints.

For example if $k^{+}\rightarrow 0$, then $k^{+}$ can approach $0$
from above and below. The pole giving us trouble stems from the
$k$-line, all other propagators are unproblematic in this case, since
their poles are at finite $k^-$. The pole in $k^{-}$ approaches
$+\infty - i\epsilon$ as one approaches $0^+$ and $-\infty+i\epsilon$
as one approaches $0^-$. This means that the $k^{-}$ pole crosses the
real axis at infinity. Hence the parton distribution is non-analytic
there.  Since the singularity is at $|k^-|=\infty$ the other
propagators have large denominators, and hence we get a zero for the
non-analytic part of integral at the breakpoint.  Thus the parton
density is continuous at the breakpoint, as claimed.  This result
enables the factorization formula to be valid in the neighborhoods of
the breakpoints.  Since the other poles in the $k^-$ integral are on
opposite sides of the real axis, the parton distribution is non-zero
at the breakpoints.

Effectively the crossover of the pole occurs when $k$ is in a
collinear-to-$B$ region, which we know is power suppressed.  This
indicates that the argument we have just given generalizes to all
graphs.

We also remark on the behavior at the endpoints.  Let us look at the
case $k^{+}\rightarrow p^{+}$. We find that another of the poles pole
runs off to $-\infty$ this time and crosses the real axis there. But
now all the other poles are on a single side of the real axis, so that
the sole contribution to the parton density comes from the pole at
infinity, and hence there is a zero of the parton density at the
endpoint.

%----------------------
\subsection{Completion of Proof}
\label{comp}

Using the definitions of the parton distributions and the hard scattering
coefficients we finally obtain Eq.\ (\ref{theorem}). Note that the theorem is
valid for the production of a real photon which directly goes into the final
state and for the production of a time-like photons that decays
into a lepton pair.

%========================================
\section{Conclusion}
\label{concl}

We have proved the factorization theorem for deeply virtual
Compton scattering up to power suppressed terms to all orders in
perturbation theory. The form of the theorem is independent of
the virtuality of the produced photon.

After this work was complete, we learned from X.-D. Ji that he
and Osborne have also constructed a proof of factorization for
DVCS \cite{Ji-new}.

%========================================
\section*{Acknowledgments}

This work was supported in part by the U.S.\ Department of Energy
under grant number DE-FG02-90ER-40577.  We would like to thank
L. Frankfurt, X.-D. Ji, A. Radyushkin, and M. Strikman for
conversations.

%========================================

\end{document}